# Energy Efficient VM Placement in a Heterogeneous Fog Computing Architecture

Abdullah M. Alqahtani, Barzan Yosuf, Sanaa H. Mohamed, Taisir E.H. El-Gorashi, and Jaafar M.H. Elmirghani, *Fellow, IEEE*
School of Electronic and Electrical Engineering, University of Leeds, LS2 9JT, United Kingdom

*Abstract*— Recent years have witnessed a remarkable development in communication and computing systems, mainly driven by the increasing demands of data and processing intensive applications such as virtual reality, M2M, connected vehicles, IoT services, to name a few. Massive amounts of data will be collected by various mobile and fixed terminals that will need to be processed in order to extract knowledge from the data. Traditionally, a centralized approach is taken for processing the collected data using large data centers connected to a core network. However, due to the scale of the Internet-connected things, transporting raw data all the way to the core network is costly in terms of the power consumption, delay, and privacy. This has compelled researchers to propose different decentralized computing paradigms such as fog computing to process collected data at the network edge close to the terminals and users. In this paper, we study, in a Passive Optical Network (PON)-based collaborative-fog computing system, the impact of the heterogeneity of the fog units' capacity and energy-efficiency on the overall energy-efficiency of the fog system. We optimized the virtual machine (VM) placement in this fog system with three fog cells and formulated the problem as a mixed integer linear programming (MILP) optimization model with the objective of minimizing the networking and processing power consumption of the fog system. The results indicate that in our proposed architecture, the processing power consumption is the crucial element to achieve energy efficient VMs placement.

## I. INTRODUCTION

The increasing popularity of time-sensitive applications such as connected vehicles, actuator networks and online gaming poses a number of challenges amongst which latency and energy efficiency have drawn researchers' attention from both academia and industry [1]. Traditionally, due to the abundance of resources available at cloud data centers, the majority of the applications are hosted by the cloud over the Internet. However, owing to the number of hops between the terminal(s) at the edge of the network and the cloud at the core, centralized processing has become costly and furthermore contributes additional latency and power consumption [2]. Thus, the trend in research has shifted to proposing a competing system capable of overcoming the aforementioned challenges [3]–[7]. Fog computing systems have been proposed as a decentralized computing infrastructure that is closer to the end-user terminal. Fog computing is formed using different resources at the edge of the network such as routers, switches, accesses points and small racks od servers, and represents a middle layer between the cloud computing and the end-user terminals layers. Accordingly, the total workload that is processed via cloud computing is reduced. However, Fog resources are limited in capacity and are highly heterogeneous in terms of computational power and energy efficiency. Consequently, resource management becomes pivotal in order to guarantee quality of service (QoS), optimum resource allocation and energy efficiency at the edge of the network [7]–[11].

The access network is an essential part of telecommunication networks that connects the end-users to their service providers. Therefore, the continuous growth in next generation of applications that require high bandwidth have led to a wide deployment of optical access networks. Besides the high bandwidth, Passive Optical Networks (PONs) are highly energy efficient due to unpowered components such as the Arrayed Waveguide grating Router (AWGR), coupler/splitter, and Fibre Bragg gratings used [12]–[16].

In a heterogeneous fog computing environment, due to equipment being manufactured by different vendors, computational resources that are placed in different locations may be of different efficiencies in terms of power consumption [17]. Several research studies investigated different optimization models to improve the performance of a heterogeneous computing system in term of data processing and power consumption [18]–[20]. The authors in [18] investigated the optimal placement of IoT applications in a heterogeneous architecture supported by edge and cloud computing. They proposed a model that utilizes a weighted objective function to optimize the latency and power consumption. In a similar study, the authors in [19] tackled the balance of power consumption and delay when optimally placing end-users requests in a three-layer heterogeneous cloudlet environment. Based on the end-users' applications' type, the appropriate cloudlet layer is selected to process the request to minimize the power consumption and delay. The authors in [20] proposed an optimization model for end-user requests' placement taking into account the processing provided by the cloud and the access network. Their proposed model utilized a weighted objective function to optimize the power consumption and delay.

Different from the aforementioned works, the rapid growth of time-sensitive applications will impose a considerable impact on the fog computing system and can degrade its performance. Therefore, we proposed a collaborative fog cells architecture in [1], [21], [22]. Our previous work in [21] maximized the Wavelength Division Multiplexing Passive Optical Network (WDM-PON) wavelength connection assignments between collaborative-fog cells to achieve full-communication between computational resources in the fog cells. Our work in [22] optimized the collaborative-fog cells' capacity by enabling the borrowing of data processing among cells to serve intensive VMs demand. Also our work in [1] investigated the VMs placements considering the inter-VMs traffic demands. We extend our previous study in this paper to consider the impact of the heterogeneity of the fog cells on the performance of the collaborative fog computing architecture. We also benefit from our previous works in energy efficiency that tackled areas such as distributed processing in the IoT/Fog layer [23]–[26], green core and data center (DC) networks [27]–[36], [37]–[42], network virtualization and service embedding in core and IoT networks [43]–[46], machine learning and network optimization for healthcare systems [47]–[50] and network coding in the core network [51], [52].

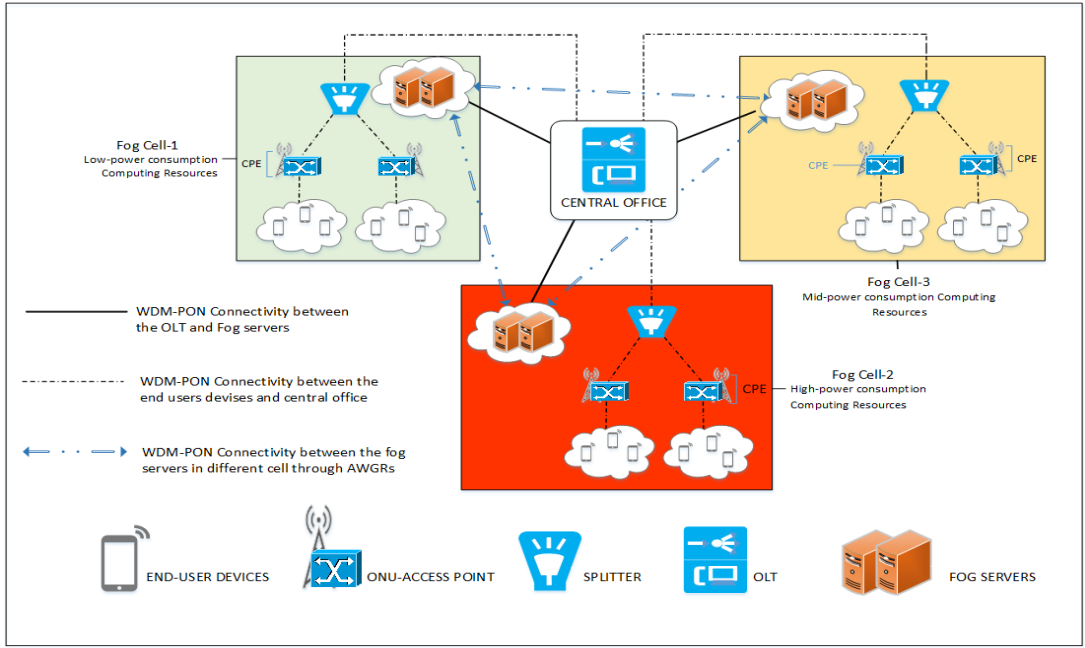

Fig. 1. The evaluated hetergenous fog computing architecture

The reminder of this paper is organized as follows: Section II introduces the proposed architecture and the optimization model. Section III presents and discusses the results. Finally, Section IV concludes the paper and outlines future work.

## II. THE OPTIMIZATION FRAMEWORK

The power consumption optimization framework in this paper considers a heterogeneous collaborative-fog computing environment that consists of three fog cells. In the following, we introduce the proposed architecture and the proposed MILP model.

### A. The Proposed Architecture

The proposed architecture, as shown in Fig. 1, is comprised of three heterogeneous collaborative-fog computing cells. Each of which is connected to the other cells using a WDM-PON, and each cell has a direct WDM-PON connectivity to an Optical Line Terminal (OLT) located in the central office. In addition, each fog cell is comprised of a networking layer and a processing layer. The networking layer deals with assembling the virtual machines requests (VMs) traffic from the end-user terminals using the Optical Networking Units (ONUs). The OLT deals with collecting data from the ONUs. The processing layer deals with processing VM requests, and is comprised of fog servers located in different cells. Each fog cell has multiple servers, where each server is equipped with an ONU device to connect to the PON network. Note that in the proposed architecture, in the interest of processing heterogeneity, we have differentiated between the servers' processing efficiencies in the fog cells.

### B. MILP Model

The proposed MILP model aims to minimize the total power consumption during the placement of the VMs. Each VM request consists of a processing demand to be performed by a CPU. This is the amount of processing required in GHz. The VM request also includes the traffic demand for communication with other VMs which is the amount of data required in Gbps, and the RAM workload which is the amount of memory in MB.

The following notations are the sets, parameters and variables used in the optimization model:

*1) Sets:*

| | |
|---|---|
| $N$ | Set of all nodes in the proposed architecture. |
| $N_m$ | Set of all neighbouring nodes to node m in the proposed architecture, $m, m \in N$. |
| $Onu$ | Set of ONUs in the fog cells, where $Onu \subset N$. |
| $Olt$ | Set of OLT in the central office, where $Olt \subset N$. |
| $S$ | Set of servers in the fog cells, where $F \subset N$. |
| $VM$ | Set of all VM request. |

*2) Parameters:*

| | |
|---|---|
| $Max_{Onu}$ | Maximum power consumption of ONUs. |
| $Idl_{Onu}$ | Idle power consumption of ONUs. |
| $Max_{Olt}$ | Maximum power consumption of OLT. |
| $Idl_{Olt}$ | Idle power consumption of OLT. |
| $Max_S$ | Maximum power consumption of server $s, s \in S$ in the processing layer. |
| $Idl_S$ | Idle power consumption of server $s, s \in S$. |
| $Cpu_S$ | CPU capacity of the server $s, s \in S$. |
| $Cpu_v$ | CPU demand of VM $v, v \in VM$. |
| $Tr_{vw}$ | Traffic Demand of between VMs $v$ and $w$, $v,w \in VM$ in Gbps. |
| $\alpha$ | A unitless factor to emphasise and de-emphasise networking power consumption. |
| $\beta$ | A unitless factor to emphasise and de-emphasise processing power consumption. |

*3) Variables:*

| | |
|---|---|
| $Tr_{sd}$ | Traffic demand between source, $s$, and destination, $d$, nodes, where $s \in N, d \in S$. |
| $Tr_{sd}^{mn}$ | Traffic flow between source node $s \in N$ and destination server node $d \in S$, traversing node $m \in N$ and $n \in Nm$. |
| $Pr_{sd}$ | Processing demand allocation between source node $s$, and destination node $d$, where $s \in N, d \in S$ |

*The MILP model is defined as follows:*

**Objective:** Minimize the weighted sum of the networking and processing power consumption of the proposed architecture [54]:

$$\alpha \times N_{PC} + \beta \times P_{PC} \quad (1)$$

The network layer's power consumption ($N_{PC}$) is comprised of the power consumption of PON's active devices in the network layer. The processing power consumption ($P_{PC}$) is comprised of the power consumption of the fog servers and ONUs attached to the servers as a communication interface. It is important to highlight that, the ONUs' power consumption profiles in the proposed model are classified into two types; (i) On/Off power profile for the ONUs in the processing layer that are attached to the servers and operate as regular transceivers and (ii) proportional profile for the ONUs that collect the demands from the end users in the access networking layer [53].

Subject to the following constraints:

$$\sum_{\substack{n \in Nm \\ m \neq n}} Tr_{sd}^{mn} - \sum_{\substack{n \in Nm \\ m \neq n}} Tr_{sd}^{nm} = \begin{cases} Tr_{sd} & m = s \\ -Tr_{sd} & m = d \\ 0 & otherwise \end{cases} \quad (2)$$

$$\forall \ s \in N, d \in S, \forall \ m \in N$$

Equation (2) is a data traffic flow conservation constraint to guarantee that the traffic demand for each VM that enters a node leaves it at the same volume except for the source and destination nodes.

$$\sum_{v \in VM} Cpu_v \, Pr_{sd} \leq Cpu_S. \quad (3)$$

$$\forall \ d,s \in S$$

Equation (3) guarantees that the processing capacity of the VMs' requests does not exceed the processing capacity of the allocated server. In addition, we used additional constraints to ensure that the capacity of links in the proposed architecture are not exceeded and constraints to ensure that all the VMs demands are met.

## III. RESULTS AND DISCUSSION

In this Section, we evaluate the energy efficiency of the proposed collaborative fog computing model. We use a weighted objective function to give preference to minimizing the processing or the networking power consumption separately and jointly. The evaluation investigates both models under three different sets of VM distributions which are 10 VMs, 15 VMs, and 20 VMs - with random distributed values for the CPU capacity requested (between 0.1 GHz and 2.6 GHz), memory capacity (between 100MB and 500MB) and data traffic (between 0.1 Gbps and 10 Gbps), as illustrated in Table 1. Three types of servers are considered in the fog cells: fog cell1 which is highlighted in green in Fig. 1 contains energy efficient servers with CPU capacity of 2.6 GHz and power consumption of 243 W maximum and 54.1 W idle. In fog cell2, highlighted in red, we considered low energy efficiency servers with CPU capacity of 2.5 GHz (power consumption of 457 W maximum and 301 W idle) and finally fog cell3, highlighted in yellow, comprised of servers with mid-range -energy efficiency with CPU capacity of 2.4 GHz (power consumption of 325 W maximum and 104 W idle). The input data used in the MILP model is shown in Table 1.

TABLE I. INPUT DATA PARAMETERS

| | |
|---|---|
| Fog Cell-1 Server's maximum power consumption (Dell PowerEdge R620) [55]. | 243 W |
| Fog Cell-1 Server's idle power consumption [55]. | 54.1 W |
| Fog Cell-1 Server's Processing capacity (CPU)[55]. | 2.6 GHz |
| Fog Cell-1 Server's Memory capacity (RAM) [55]. | 24 GB |
| Fog Cell-2 Server's maximum power consumption (Dell PowerEdge R740)[56]. | 457 W |
| Fog Cell-2 Server's idle power consumption[56]. | 301 W |
| Fog Cell-2 Server's Processing capacity [56] | 2.5 GHz |
| Fog Cell-2 Server's Memory capacity (RAM)[56] | 16 GB |
| Fog Cell-3 Server's maximum power consumption Hitachi, Ltd. HA8000/RS220-hHM)[57]. | 325 W |
| Fog Cell-3 Server's idle power consumption[57]. | 104 W |
| Fog Cell-3 Server's Processing capacity [57] | 2.4 GHz |
| Fog Cell-3 Server's Processing capacity [57] | 32 GB |
| Memory capacity (RAM) of the VMs [53]. | 100MB - 500MB |
| OLT Maximum power consumption [58]. | 1940 W |
| OLT idle power consumption | 1746 W |
| OLT data rate [58]. | 8600 Gbps |
| ONU Maximum power consumption [59]. | 2.5 W |
| ONU idle power consumption[59]. | 1.5 W |
| ONU data rat e[59]. | 10 Gbps |
| VMs Traffic Demands [53]. | 1Gbps–5 Gbps |
| Capacity of Optical physical link [60]. | 32 wavelengths 40 Gbps per wavelength |

We have evaluated the total power consumption of the collaborative fog architecture under two scenarios: 1) $\alpha \gg \beta$, and 2) $\alpha = \beta$. In the first scenario, we give preference to optimizing the network power consumption only whilst in the second scenario we give the two weights equal value so that both processing and networking power consumption are equally important. In Fig. 2 and Fig. 3 we present the total power consumption and its breakdown into the networking power consumption and processing power consumption against the number of VMs. In the second scenario, the proposed model reduced the total power consumption of placing 10 VMs, 15 VMs and 20 VMs by 45%, 28% and 23%, respectively in comparison to the model in scenario 1. This is due to the fact that the processing power consumption contribution is much greater than that of the network (due to the higher power consumption of servers compared to networking equipment), hence VMs can be better allocated to more efficient servers when $\alpha \gg \beta$. However, the network power consumption remains constant due to the passive nature of the AWGR-based PON connecting the fog cells and its low power consumption.

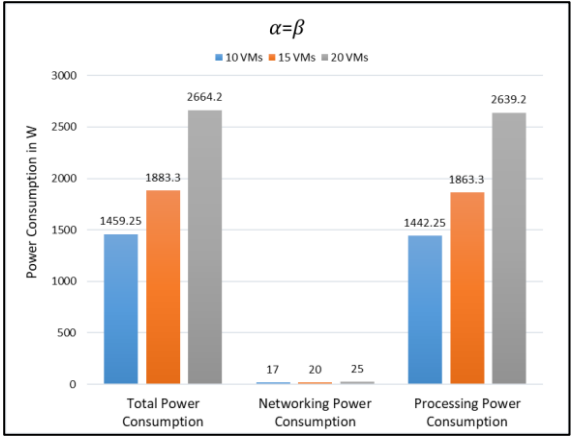

Fig. 2. The Total Power Consumption when $\alpha = \beta$

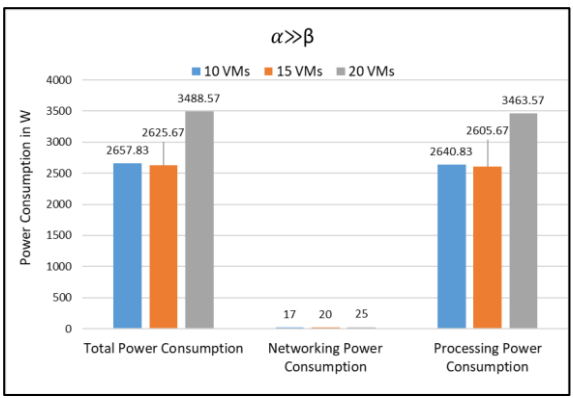

Fig. 3. The total power consumption when $\alpha \gg \beta$.

Fig. 4 and Fig. 5 present the total power consumption distribution while, Fig. 6 and Fig. 7 present the server utilization against different VM workloads. It can be clearly seen that when $\alpha = \beta$, the model favours fog cell 1 during the VM allocation, due to the processing efficiency of the servers in that cell. As was expected, the VMs were only allocated to the servers in fog cell 2 due to capacity limitations at 20 VMs test case. On the other hand, when $\alpha \gg \beta$, the model emphasizes the networking power consumption. Therefore, the selection of the energy efficient processing resources is not a consideration in this case.

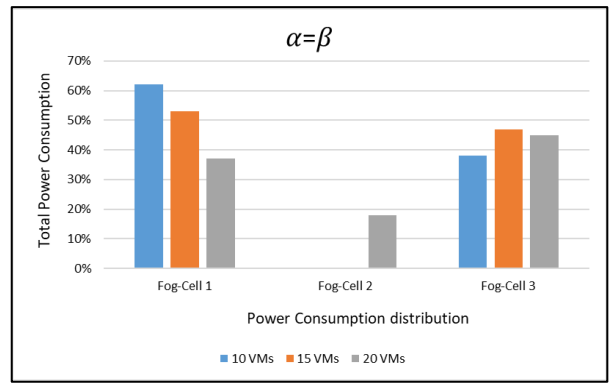

Fig. 4. Total power consumption distribution when $\alpha = \beta$.

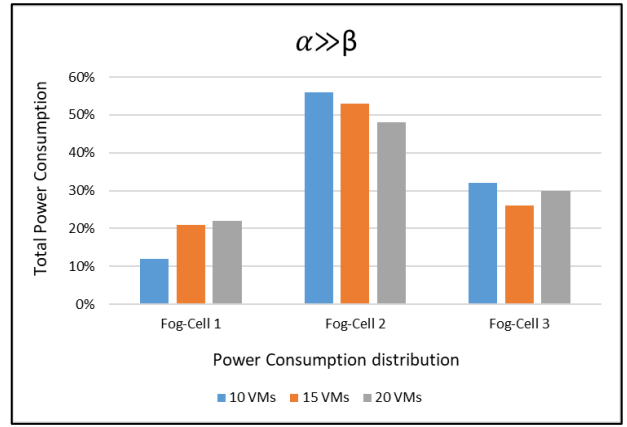

Fig. 5. Total power consumption distribution when $\alpha \gg \beta$.

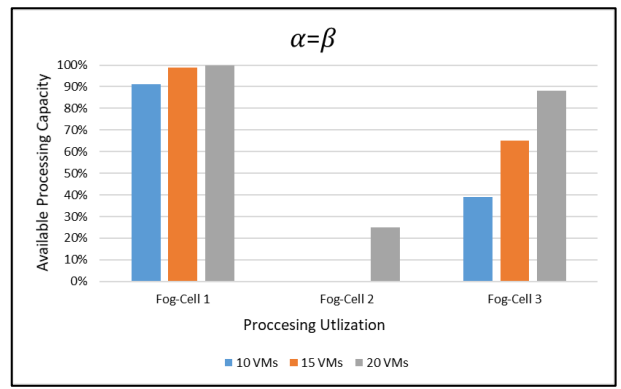

Fig. 6. The processing utlization when $\alpha = \beta$.

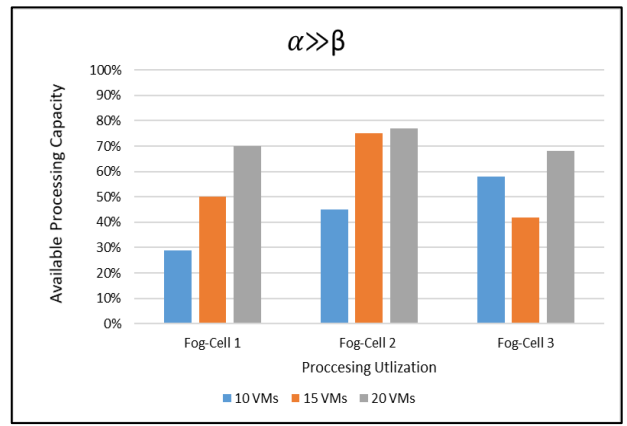

Fig. 7. The processing utilization when $\alpha = \beta$.

## IV. CONCLUSIONS

In this paper, we studied the VM placement problem in a heterogenous fog environment with collaborative fog cells interconnected with a PON. We proposed a weighted objective function whose goal was to study the impact of optimizing processing and networking separately. The results showed that VMs were better placed when the appropriate weight was set to emphasize the processing power consumption. The networking power consumption impact is minimal due to the passive nature of PON. Our results show that the total power consumption of placing 10 VMs, 15 VMs and 20 VMs can be reduced by 45%, 28% and 23% in the scenarios considered. Future work includes extensions of the current optimization model to consider delay, and mobility-aware VM placement.

## ACKNOWLEDGMENT

The authors would like to acknowledge funding from the Engineering and Physical Sciences Research Council (EPSRC) INTERNET (EP/H040536/1), STAR (EP/K016873/1) and TOWS (EP/S016570/1) projects. The first author would like to acknowledge the Government of Saudi Arabia and JAZAN University for funding his PhD scholarship.